\documentclass[10pt]{article}

\usepackage[letterpaper, margin=1in]{geometry}

\usepackage{cite}
\usepackage{amsmath,amssymb,amsfonts}
\usepackage{graphicx}
\usepackage{textcomp}
\usepackage{xcolor}
\usepackage{booktabs}
\usepackage{array}
\usepackage{url}

\def\BibTeX{{\rm B\kern-.05em{\sc i\kern-.025em b}\kern-.08em
    T\kern-.1667em\lower.7ex\hbox{E}\kern-.125emX}}

\newcommand\gk[1]{{\color{black}{#1}}}
\begin{document}

\title{Performance and Cost-Aware Cache Provisioning}
\author{Ridwanul Tanvir and George Kesidis\\
Computer Science \& Engineering Dept, Electrical Engineering Dept\\
Pennsylvania State University\\
University Park, PA, 16802, USA\\
\{rpt5409,gik2\}@psu.e.du
}

\date{}
\maketitle

\begin{abstract}
While traditional cache policy evaluations fix capacity—often at 0.1\% of the dataset—and measure the resulting hit rate, practical edge-cloud deployments require balancing both storage and computational overhead as billed resources. Consequently, system operators frequently focus on a different objective: determining the minimum cache size needed to satisfy a specific Service-Level Objective  (SLO) hit-rate. This paper explores this SLO-centric paradigm by analyzing the minimum capacity and execution time each policy requires to hit a defined target. Additionally, we show that dynamically adjusting the segment ratio in segmented policies based on historical workload patterns enhances efficiency. Through evaluations across real-world and synthetic traces, we present a novel hybrid segmented policy that reduces capacity requirements while keeping processing costs low.
\end{abstract}

\section{Introduction}
\label{sec:introduction}

Modern cache deployments must balance service performance against the resources
reserved to obtain that performance. Production cache engines expose this
tradeoff directly: memory, flash space, and computation to implement cache replacement (eviction) policies
are provisioned resources, while applications impose performance requirements on latency,
throughput, or hit rate \cite{CacheLib20}. Related work has also studied
SLO-driven cache sizing, dynamic cache reallocation, and cache sharing
(including object apportionment) across
services \cite{zhang2021redy,berger2018robinhood,Stoica16,Towsley17}.
Complementary caching problems include online service placement,
distributed data access, content placement in cache networks,
dependency-aware caching, caching of dynamic content, and semantic
response caching
\cite{FanHou23,ZuoTang24,ZhangYeh24,IoannidisYeh16,
DallotSchmid24,QuanLin26,SemanticCache26}.
The conventional evaluation question asks, for a fixed cache size, which replacement
policy obtains the highest hit rate \cite{Krunz04,HMK23,YZQ23,ZYY24}? That view
is important, but it does not directly answer a provisioning question faced by
an operator (user): for a required target hit rate (user Service-Level Objective (SLO)), 
how much cache capacity does each
policy need (what is its cost)? We study this SLO-aware viewpoint by treating the target object 
hit-rate as an explicit requirement and comparing policies by the minimum resident
cache capacity needed to satisfy it.

Content caches are typically located at edge or near-edge cloud platforms whose
resources are much more expensive than in remote cloud platforms housing the origin
servers of the content.
Reducing required cache capacity has benefits beyond a single billing model. On
private or shared servers, a smaller cache leaves more DRAM, SSD space, or disk
capacity for colocated applications and other services. In cloud deployments, a
lower memory or storage requirement may also permit a smaller virtual-machine
configuration, a lower storage allocation, or a less expensive caching tier.
These two benefits are related but distinct: one is resource availability for
other work, and the other is monetary cost in an elastic deployment.

The pricing implication must be interpreted carefully. Cloud resources are often
allocated in coarse units, and a small reduction in cache capacity does not
necessarily produce an immediate proportional cost reduction. The granularity of
allocation, the frequency with which resources can be resized or released, and
the placement of the cache all matter. Remote cloud sites may offer large
resource pools but coarse instance boundaries. Content caches often reside at
the network edge or near edge.
Near-edge deployments may offer
more capacity than the edge with higher access latency, and edge deployments
may have more limited and costlier 
memory, storage, and compute resources with finer granularity. 
Thus, edge and near-edge
systems motivate careful resource accounting, while production traces with large
working sets may not be able to achieve a high hit rate using 
relatively small amounts of edge-cloud resources, particularly memory.

Let \(B\) denote resident cache capacity and let \(E\) denote computational
effort such as CPU time, metadata processing, or history-management overhead.
For a target hit rate \(\eta\) (the operator's/user's performance SLO), an idealized SLO aware objective is
\begin{equation}
\min_{B,E} P(B,E)
\qquad
\text{subject to}
\qquad
H(B)\geq \eta,
\label{eq:intro-cost-objective}
\end{equation}
where \(P(B,E)\) denotes a deployment-specific resource loss cost SLO over storage
and computation that is specified by and ``internal to" the operator/user. In this paper, 
we simply take \(P=B\), the resident cache capacity.
Under this objective, a policy that reaches the same performance SLO with less resident
capacity, less computation, or both is preferable under the resource-aware view.
We explain at the end of Section \ref{sec:workloads} below how problem \ref{eq:intro-cost-objective} is a kind of 
inverse of that addressed by the working-set approximation for caches \cite{DS72}.

Computation is part of this cost even when end-to-end latency is dominated by
(and can be pipelined with)
memory, disk, or network I/O. Replacement policies differ in the amount of
metadata they maintain, the work performed on hits, and the work needed to
choose victims to evict on misses. 
These differences matter for shared servers and are
especially relevant in constrained deployments where both memory and compute are
valuable. Recent production-oriented cache designs explicitly account for
hit-path metadata cost and implementation complexity
\cite{CacheLib20,Segcache21,YZQ23,ZYY24}. We therefore report not only capacity requirements but also execution
cost when comparing policies.

Workloads also change over time. Web and object-cache workloads are often skewed,
heavy-tailed, and sensitive to changes in object popularity and reuse
\cite{Shenker99,Krunz04,Atikoglu12,YangTwitter20,CacheLib20,META-traces}. A fixed cache-policy parameter
may therefore perform well during one interval and poorly during another. This
paper focuses on adapting the internal segment ratio \(r\) of segmented caching
policies while keeping the total cache capacity fixed. The adaptation is causal:
each decision uses only observations from completed request blocks and does not
use future requests, future reuse distances, offline phase labels, or
counterfactual cache states. A study of dynamic adaptation of the total cache capacity
\(B\) is left for future work.

\textbf{Contributions.} This paper makes five contributions. First, it presents
an SLO-aware evaluation methodology that compares policies by the minimum cache
capacity required to achieve a target hit rate. Second, it applies this
methodology to multiple caching-policy families under a common experimental
framework. Third, it introduces GAMP, a segmented caching policy designed to
obtain strong hit-rate performance with low required cache capacity and
practical computational overhead for a range of SLOs and workloads. Fourth, it studies
causal adaptation of the segment ratio \(r\) using past workload observations.
Fifth, it evaluates both resident-capacity requirements and policy execution
cost, reflecting that deployed caches consume storage and computation.


\textbf{This paper is organized as follows.} Section~\ref{sec:background}
introduces the cache model, policy families, and GAMP. Section~\ref{sec:experimental-setup} explains workloads, and SLO-aware
evaluation methodology. Section~\ref{sec:dynamic-r} studies segmented caching,
the need to adapt \(r\), and causal \(r\)-adaptation, together with the
main experimental comparisons of minimum cache size and computational overhead in Section~\ref{sec:experimental-results}.
In Section \ref{sec:practical-provisioning}, we discuss practical 
memory pricing issues in edge clouds.
The final section concludes and discusses future work.

\section{Background, Policy Families, and Evaluation Methodology}
\label{sec:background}

A content cache stores data objects so that later requests can be served
without accessing a slower backend such as a remote service, lower cache tier,
disk, or origin store. Hierarchical web-cache models, production cache engines,
and cache-replacement surveys describe this common structure across deployment
settings \cite{Che02,Krunz04,CacheLib20,HMK23}. Practical caching engines
maintain a cached-object store, an index, replacement-policy metadata, and
allocation machinery \cite{CacheLib20}. A request is a \emph{hit} if the object
is resident when the request arrives and a \emph{miss} otherwise. On a miss, an
admission rule decides whether the fetched object enters the cache; if the cache
is full, the replacement policy selects one or more resident victims.

The resident payload is distinct from policy metadata. A \emph{resident entry}
contains the cached object and the metadata needed to manage it. A
\emph{ghost entry} contains only an identifier for a recently evicted object.
A request that matches a ghost entry is still a miss because the payload is not
resident. The match is nevertheless useful because it provides causal evidence
that an evicted object returned and may deserve more protection if it is
admitted again. Throughout this paper, ghost queues do not count toward
resident capacity \(B\) \cite{JohnsonShasha94,Megiddo03,YZQ23}.

Many policies divide resident capacity into a probationary region and a
protected region. The probationary, or cold, region receives newly admitted
objects and filters objects that do not demonstrate reuse. The protected, or
hot, region stores objects with stronger reuse evidence. The terms cold and
hot describe functional roles only; policies differ in the events that move an
object from cold to hot and in whether ghost entries influence that movement
\cite{Karedla94,JiangZhang02,JohnsonShasha94,Megiddo03,YZQ23}.
Prior work has also studied cache admission, reuse prediction, learned
replacement, workload-aware eviction, and scalable cache implementation
\cite{BergerSH17,BeckmannCC18,EinzigerFM17,SongBLL20,RodriguezYLP21,YangMYR23}.
Related studies examine caching in key-value service, edge, file-system, and
flash-backed storage settings
\cite{ChenZZCZ25,PanWCL22,ZhangTLHJL22,XiaoLRWZ24}.

\subsection{Policy Families}

The evaluated policies fall into three families. A (not segmented) \emph{pure} policy applies
one replacement rule to the entire resident cache and has no resident partition
parameter.
FIFO maintains insertion order and evicts the oldest resident irrespective of use, 
i.e., a hit does not refresh the object or change its position (eviction rank). This makes
FIFO inexpensive on the hit path, but it can evict frequently reused objects if
they happen to be old. LRU instead orders objects by most recent access and
evicts the Least Recently Used object. It is related to the classical
working-set view of locality, but every hit must update recency metadata, which
can create synchronization and memory-access overhead in high-throughput caches
\cite{DS72,Towsley90,Krunz04,YZQ23,ZYY24}.

SIEVE is also a pure policy, but it avoids LRU's hit-path reordering
\cite{ZYY24}. It keeps cached objects in insertion order, attaches one
visited bit to each cached object, and maintains a persistent eviction hand.
A hit sets the visited bit and leaves the object in place. When eviction is
needed, the hand scans from older objects toward newer objects. If the hand
finds a visited object, it clears the bit and gives the object another chance;
the first unvisited object encountered is evicted. Thus SIEVE records reuse on
the hit path with a bit write, and performs the retention decision later on the
eviction path. This gives reused objects lazy protection while allowing objects
with no observed reuse to be removed quickly.

A \emph{segmented-pure} policy divides resident capacity among regions with
different roles. SLRU uses multiple LRU segments: new objects enter a cold
segment, hits move objects toward more protected segments, and evictions are
taken from the coldest nonempty segment \cite{Karedla94}. 2Q separates
short-term recency from longer-term reuse using a resident FIFO cold queue, a
nonresident ghost queue, and an LRU hot queue \cite{JohnsonShasha94}. A new
object enters the cold queue. A hit while it is still resident in cold does not
move it to hot; only after the object leaves cold and later returns through the
ghost history is it admitted into the hot queue. This rule prevents a short
burst or single correlated reference from immediately receiving long-term
protection.

ARC is a segmented LRU policy with online adaptation \cite{Megiddo03}. It keeps
two resident LRU lists: one for recently seen objects and one for objects with
stronger frequency evidence. It also keeps two nonresident ghost lists recording
objects evicted from those resident lists. A hit in the recency ghost list
indicates that the recency side was too small, while a hit in the frequency
ghost list indicates that the frequency side was too small. ARC uses those
ghost hits to adjust the target split between its two resident lists, so the
policy can shift capacity between recency and frequency without seeing future
requests.

S3-FIFO is a segmented FIFO--FIFO policy designed to retain FIFO's low
metadata-update cost while filtering one-hit objects \cite{YZQ23}. It has a
small resident FIFO cold queue, a larger resident FIFO hot queue, and a
nonresident FIFO ghost queue. Newly admitted objects enter the cold queue with
a small counter. Hits increment the counter but do not change FIFO order. When
an object reaches the tail of cold, it moves to hot only if its counter shows
sufficient reuse; otherwise the payload is evicted and the identifier enters
the ghost queue. A later request for a ghosted object is still a miss, but the
ghost hit admits the returning object directly to hot. In the hot queue,
counter-based reinsertion gives reused objects additional protection while
preserving FIFO-style ordering.

A \emph{segmented-blended} policy keeps the parent policy's admission and movement rules
but replaces one resident replacement primitive with another. Prior work uses
this composition to construct segmented policies such as ARC-SIEVE and 2Q-SIEVE by
replacing selected LRU components with SIEVE \cite{ZYY24}. ARC-SIEVE retains
ARC's resident cold list, ghost histories, and adaptive target update, but uses
SIEVE for replacement in the hot region. 2Q-SIEVE retains 2Q's FIFO cold queue
and ghost-hit admission rule, but uses SIEVE instead of LRU for the hot queue.
S3-FIFO-SIEVE follows the same idea for S3-FIFO: it keeps S3-FIFO's cold FIFO
filter and ghost admission behavior, but manages the hot segment with SIEVE
rather than FIFO.

\subsection{Ghost-Aware Mode-based Promotion (GAMP)}

GAMP is the proposed segmented-blended policy in
this paper. Its resident structure is FIFO--SIEVE: newly admitted objects enter
a FIFO cold segment, and protected objects are stored in a SIEVE-managed hot
segment. If \(r\) is the resident fraction assigned to cold, then cold and hot segment sizes are
\begin{equation}
|C|_{\max}=\lceil rB\rceil,\qquad |H|_{\max}=B-|C|_{\max}.
\label{eq:gamp-partition}
\end{equation}
GAMP also maintains metadata-only ghost
histories for objects evicted from resident segments. These ghost histories
cannot serve requests and are not counted as resident capacity.

GAMP differs from a simple FIFO--SIEVE blend in its promotion rule. In the
initial conservative mode, a missed object enters cold unless it returns
through a ghost history, in which case it is admitted directly to hot. A
resident hit in cold increments a small reuse counter but does not immediately
move the object. When the object reaches the FIFO tail of cold, GAMP promotes
it to hot only if the counter meets the current promotion threshold; otherwise
the payload is evicted and the object identifier enters a ghost history. Hits
in hot are handled by SIEVE's visited-bit mechanism, so hot objects receive
lazy protection without LRU-style movement on every hit.

GAMP's mode-based behavior changes how aggressively objects are promoted or
admitted to hot, but the signals remain causal. The policy uses observations
already seen in the request stream, including resident hits, evictions, and
ghost returns. It does not use future requests, future reuse distances, phase
labels, or offline replay results.

This design separates three concerns. FIFO cold admission provides an
inexpensive filter for newly observed objects. SIEVE hot replacement reduces
hit-path reordering overhead while still protecting objects with observed
reuse. Ghost-aware modes use nonresident return evidence to decide when the
policy should become more aggressive about admitting or promoting objects to
hot. Section~\ref{sec:dynamic-r} evaluates how this policy family interacts
with the SLO-aware minimum-capacity objective.

\begin{figure}[!t]
\centering
\includegraphics[width=\linewidth]
{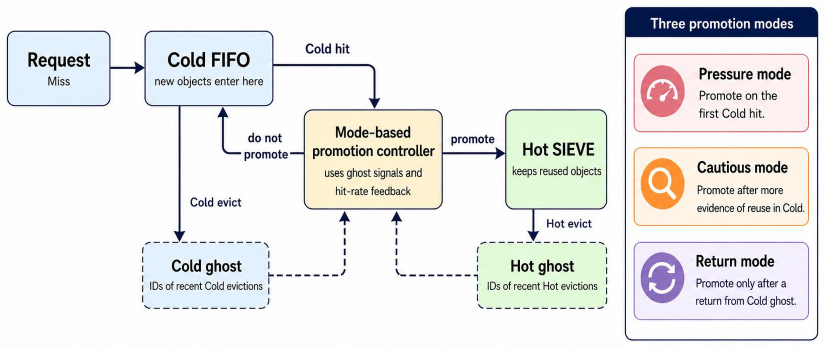}
\caption{GAMP architecture. Requests enter a FIFO cold segment before either
being evicted into metadata-only ghost histories or promoted into a SIEVE-managed
hot segment. The mode-based promotion controller uses causal signals from cold
hits, ghost returns, and hit-rate feedback.
Ghost entries contain identifiers only and do not count toward resident capacity \(B\).
}
\label{fig:gamp-architecture}
\end{figure}

\begin{table}[t]
\centering
\caption{Policy roles used in the evaluation. Ghost entries contain
identifiers only and do not count toward resident capacity \(B\).}
\label{tab:policy-summary}
\scriptsize
\setlength{\tabcolsep}{3.1pt}
\renewcommand{\arraystretch}{1.08}
\begin{tabular}{p{0.15\linewidth} p{0.38\linewidth} p{0.35\linewidth}}
\hline
Policy & Resident organization & Reuse or adaptation signal \\
\hline
FIFO & One FIFO queue & Hits do not change order \\
LRU & One LRU list & Hits refresh recency \\
SIEVE & One queue with visited bits & Eviction hand gives lazy protection \\
SLRU & Multiple LRU segments & Hits move objects toward protected segments \\
2Q & FIFO cold, LRU hot & Cold ghost hit admits to hot \\
ARC & LRU cold, LRU hot & Ghost hits adapt the target split \\
S3-FIFO & FIFO cold, FIFO hot & Counters and ghost hits move objects to hot \\
S3-FIFO-SIEVE & FIFO cold, SIEVE hot & S3-FIFO admission with SIEVE hot eviction \\
GAMP & FIFO cold, SIEVE hot & Ghost-aware modes adjust promotion behavior \\
\hline
\end{tabular}
\end{table}

\section{Experimental Setup}
\label{sec:experimental-setup}

\subsection{Workloads} \label{sec:workloads}

\gk{We use three production traces from Meta
\cite{META-traces}. Each trace is processed in chronological order, and Cache
Capacity is herein reported in objects. Results for versions of
considered cache replacement policies for variable-length objects
are omitted here for lack of space.
}

\begin{table}[t]
  \centering
  \caption{Summary of the traces used in the experiments.}
  \label{tab:trace_summary}
  \small
  \begin{tabular}{@{}lcc@{}}
    \toprule
    \textbf{Trace} & \textbf{\# Objects (N)} & \textbf{\# Requests} \\
    \midrule
    META Trace 1       & 12.26 million & 45.62 million \\
    META Trace 2       & 35.04 million & 96.68 million \\
    META Trace 3       & 29.03 million & 88.47 million \\
    Synthetic Trace 1  & 100,000       & 10.00 million \\
    Synthetic Trace 2  & 1.00 million  & 10.00 million \\
    \bottomrule
  \end{tabular}
\end{table}

We also use two controlled synthetic Poisson traces, with independent
object selection for each query,
to isolate nonstationary popularity. The two synthetic traces each consist of three Zipf phases, with object popularities sampled according to a Zipf distribution. The Zipf exponent changes across the phases as follows:
\begin{equation}
\alpha:\quad 1.6 \rightarrow 0.6 \rightarrow 1.6 .
\label{eq}
\end{equation}

The high-skew phases concentrate requests on a smaller hot set, while the
low-skew phase spreads requests across a broader part of the catalog.
Because
the catalog remains fixed, these workloads test adaptation to changing
popularity rather than changing key space \cite{Shenker99} 
\gk{seen in the production traces.}

Note that a Poisson
query process with the objective of each query independently selected from
the \emph{same} object-popularity distribution (so the process
is stationary) is a.k.a. the Independent Reference Model (IRM). 
Under the IRM for the pure LRU policy, 
a  ``working-set" approximation 
can be used to numerically estimate the cache-hit probability $h_i$
for each object $i$ \emph{given} the cache capacity $B$ \cite{DS72}. 
The average hit probability can then
be estimated as $H=\sum_{i=1}^N h_i v_i$ where $v_i > 0$ is the  (e.g., Zipf)
popularity of object $i$
(of $N$ total objects) normalized so that $\sum_{i=1}^N v_i =1$.
(More recent work extends working-set approximations to networks of caches, e.g., 
\cite{Che02,Towsley10}, where we note that caches with segmented policies can be viewed as tandem caches.)
So, problem \eqref{eq:intro-cost-objective} is the inverse of the working-set approximation: 
For LRU, \emph{given} $H=\eta$,\footnote{When object $i$ has length 
$\ell_i$ bytes, then $B\approx \sum_{i=1}^N \ell_i h_i$ bytes, where 
the hit probability of object $i$ is
$h_i = 1-\exp(-x_i t)$,  $x_i\propto v_i$ is the mean query rate of object $i$, and (in an asymptotic regime, 
common eviction time) $t>0$ solves
$N \eta = \sum_{i=1}^N h_i$.}
$$ B\approx N \eta ~~\text{[objects]}.$$

\subsection{SLO-Aware Evaluation}

Let \(\pi\) be a cache replacement policy, \(\eta\in(0,1)\) be a target hit rate, and
\(I_i\in\{0,1\}\) indicate whether request \(i\) is a hit. The full-trace hit
rate is
\begin{equation}
H_{\pi}=\frac{1}{N}\sum_{i=1}^{N} I_i .
\label{eq:hit-rate}
\end{equation}

A run satisfies the SLO only if \(H_{\pi}\geq\eta\). We include the warm-up
period in all reported results: every run starts from an empty cache and is
measured from the first request through the end of the trace.

For each policy--trace--target combination, the offline minimum fixed-capacity
baseline is
\begin{equation}
B^{\star}_{\pi,\eta}
=
\min\{B\in\mathbb{N}:H_{\pi}(B)\geq\eta\}.
\label{eq:minimum-fixed-capacity}
\end{equation}
For policies with an explicitly studied partition parameter, the search also
optimizes over the declared fixed parameter grid. This baseline is offline and
non-causal because it uses repeated full-trace replays to find the smallest
feasible constant capacity. It is not an online deployment method; it is a
reference against which causal adaptive methods can be compared.

The adaptive experiments process the trace once in chronological order using a
continuous cache instance. The controller may update its control variables only
after observing a completed request block. Decisions for the next block may use
past requests, hits, misses, admissions, evictions, and current cache state, but
may not use future requests, future block statistics, phase labels, or
counterfactual replays. This causal boundary is central to the paper.

\section{Dynamic Adaptation of Cache-Policy Hyperparameters}
\label{sec:dynamic-r}

This section develops a causal rule for adapting a cache-policy segment ratio $r$
while keeping the total Cache Capacity $B$ fixed. The rule is not tied to one
replacement policy. It applies to any segmented cache design that exposes a
probationary region, a protected region, and past-only feedback about whether
the current split is filtering too aggressively or reserving too much space for
new objects. S3-FIFO is used below as a motivating instantiation because its
cold fraction is explicit and its ghost history provides a direct causal signal;
the controller itself is a general segment-ratio adaptation mechanism.

\begin{figure}[!t]
\centering
\includegraphics[width=\linewidth]
{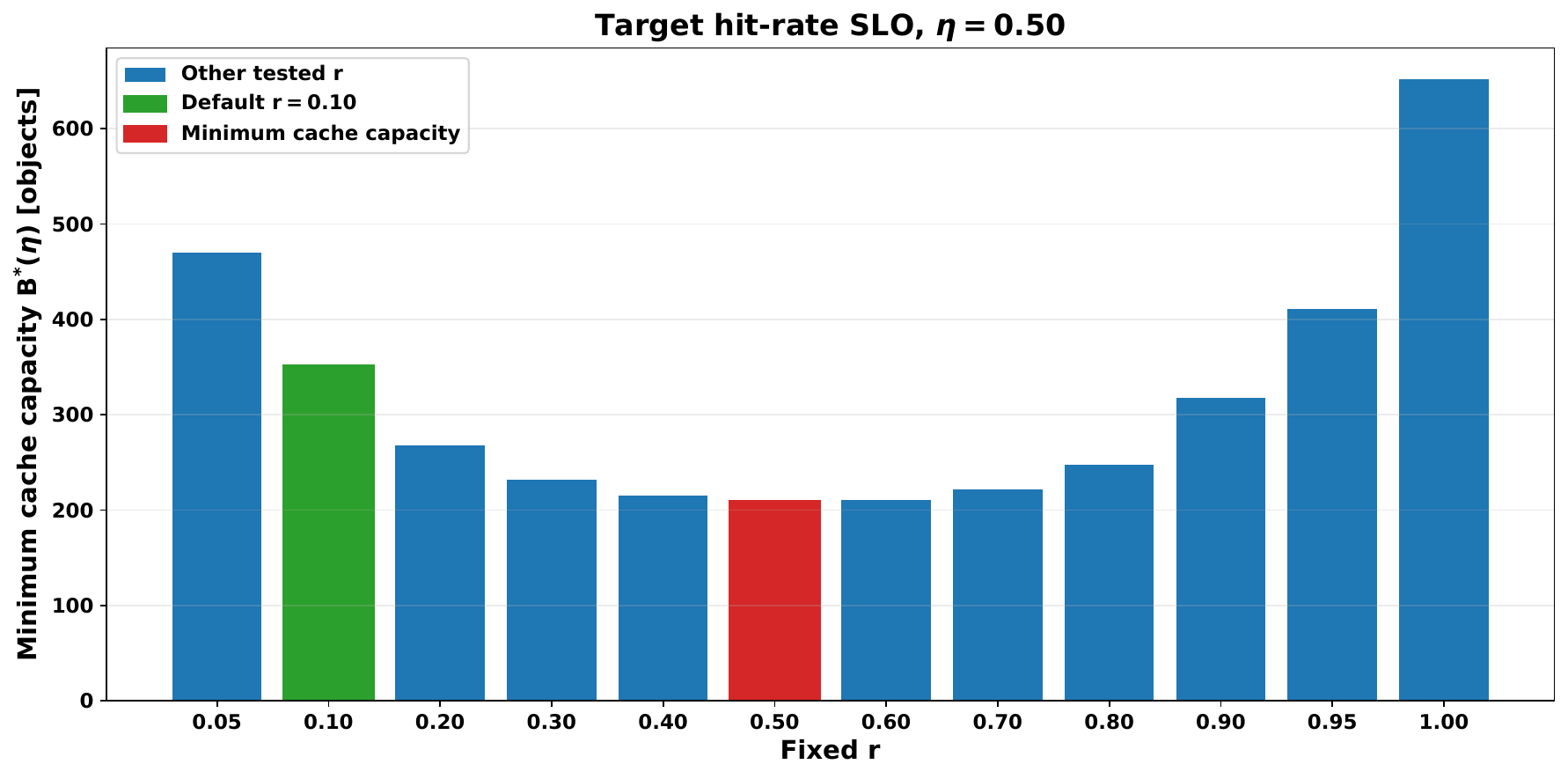}
\caption{Fixed-\(r\) sensitivity of S3-FIFO on META Trace 1 for a
representative target hit-rate SLO. Each bar reports the minimum Cache Capacity \(B\),
counted in objects, required to satisfy the same hit-rate SLO for one fixed
cold-segment fraction \(r\). The visible spread across fixed partitions
motivates an online controller that adapts \(r\) using only past observations
from the chronological request stream.}
\label{fig:representative-fixed-r}
\end{figure}

\begin{figure*}[!t]
\centering
\includegraphics[width=0.82\textwidth]
{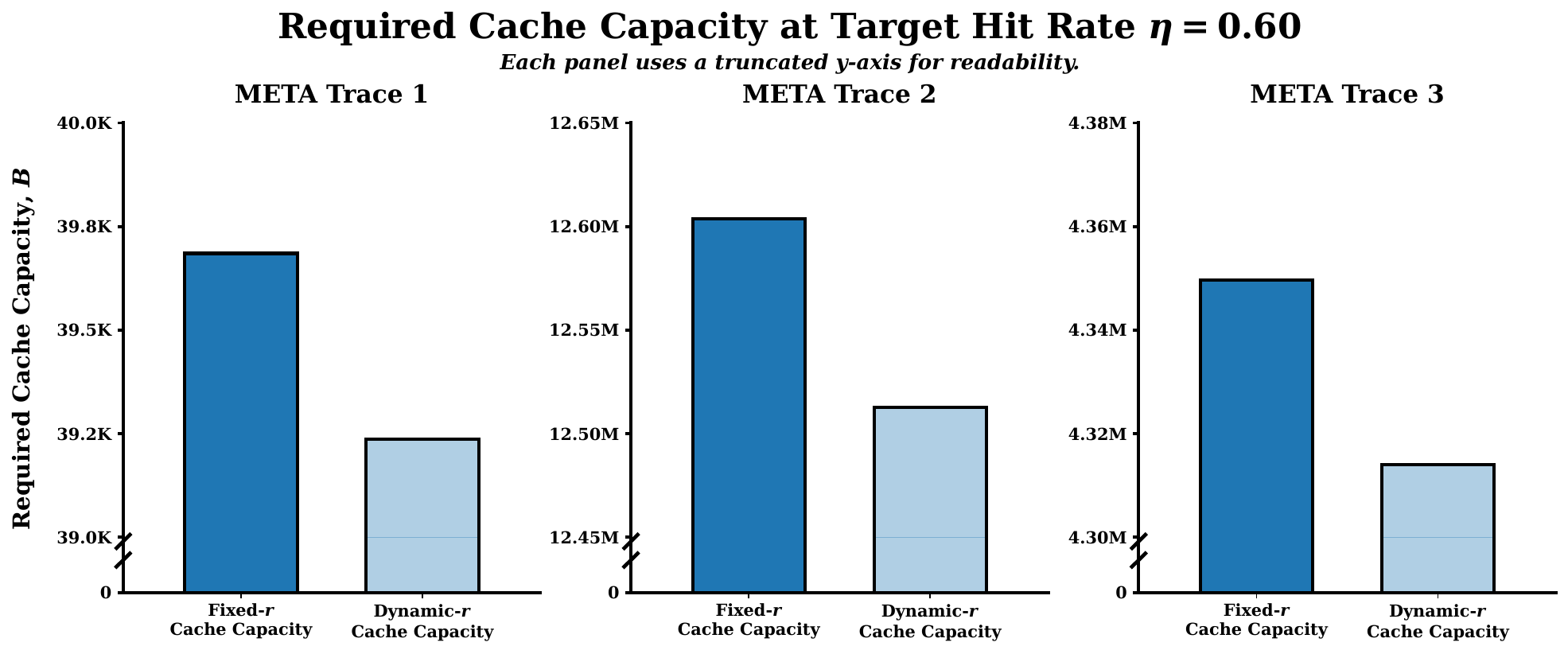}
\caption{Required Cache Capacity for a representative high target hit-rate SLO
under the best fixed-\(r\) baseline and the dynamic-\(r\) controller on three
META traces. Cache Capacity is counted in objects; each replay keeps the
total capacity \(B\) fixed while the dynamic controller adapts only the internal
segment ratio \(r\). Each panel uses a truncated vertical axis for readability.}
\label{fig:dynamic-r-required-capacity-eta060}
\end{figure*}

\subsection{Motivating Sensitivity of the Segment Ratio}

Consider a segmented policy \(\pi\) with fixed Cache Capacity \(B\), a
probationary/cold segment \(C\), and a protected/hot segment \(H\). The segment ratio
\(r\) sets
\begin{equation}
|C|_{\max}=\lceil rB\rceil,
\qquad
|H|_{\max}=B-|C|_{\max}.
\label{eq:dynamic-r-segment-split}
\end{equation}
A larger \(r\) gives newly admitted objects more time in \(C\) to demonstrate
reuse before eviction. A smaller \(r\) reserves more of the same fixed Cache Capacity
for objects that have already accumulated reuse evidence. Thus \(r\) is an
internal capacity allocation decision, not a change to the total cache size.

Fig.~\ref{fig:representative-fixed-r} uses S3-FIFO only as a motivating example
of this broader issue. The minimum capacity needed to reach the same hit-rate
target can vary substantially with \(r\). On META Trace 1, moving away from the
default split can substantially reduce the capacity needed to meet the same
SLO. For any segmented
policy \(\pi\) and fixed partition \(r\), the reference quantity is
\begin{equation}
B_{\min}^{\pi}(r,\eta)=
\min\{B\in\mathbb{N}:H_{\pi}(B,r)\geq\eta\}.
\label{eq:fixed-r-minimum}
\end{equation}
The best tested value \(r^\star\) is selected by minimizing this quantity over
the tested grid. This fixed-\(r\) sweep is non-causal because it uses repeated
full-trace replay to choose \(r^\star\). The goal of dynamic \(r\) is to replace
this offline parameter selection with an online rule that changes the split
during a single chronological pass.

\subsection{A Policy-Agnostic Dynamic-\(r\) Rule}

The controller keeps total \(B\) fixed and updates only the allocation between
\(C\) and \(H\) after each chronological request block. The block length is fixed
before the experiment, so the update frequency is not chosen using future trace
behavior. During block \(k\), the policy runs normally and records past-only
counters. Let \(M_k\) be the number of cache misses, \(G_k\) the number of
metadata-history matches, \(A_k^C\) the number of admissions to \(C\), \(D_k^C\)
the number of objects that leave \(C\) without promotion to \(H\), \(R_k^H\) the
number of cache hits served by \(H\), \(R_k\) the total number of cache hits, and
\(L_k\) the number of requests in the block.

These counters form two directional pressures. A history match means that an
object removed from \(C\) was requested again while its identifier remained in
metadata, suggesting that \(C\) may be too small. The raw pressure to increase
\(r\) is
\begin{equation}
P_k^{\mathrm{raw}}=\frac{G_k}{M_k+1},
\label{eq:dynamic-r-increase-raw}
\end{equation}
where the added one handles blocks with no misses. Evidence for decreasing
\(r\) comes from \(C\) occupancy that does not lead to promotion while \(H\)
serves most cache hits:
\begin{align}
U_k&=\frac{D_k^C}{A_k^C+1},
&
Q_k&=\frac{R_k^H}{R_k+1},
&
W_k^{\mathrm{raw}}&=U_kQ_k .
\label{eq:dynamic-r-decrease-raw}
\end{align}
A larger \(W_k^{\mathrm{raw}}\) indicates that newly admitted objects are
consuming space in \(C\) without observed reuse, while \(H\) is responsible for
most useful cache service. The controller smooths the two directional signals
with first-order EWMAs,
\begin{align}
I_k &= \alpha P_k^{\mathrm{raw}} + (1-\alpha)I_{k-1},\\
D_k &= \alpha W_k^{\mathrm{raw}} + (1-\alpha)D_{k-1},
\label{eq:dynamic-r-smoothed-evidence}
\end{align}
and uses their difference
\begin{equation}
E_k=I_k-D_k
\label{eq:dynamic-r-net-evidence}
\end{equation}
as the signed evidence for changing the split. Positive evidence indicates that
\(C\) should receive more of \(B\); negative evidence indicates that \(H\) should
receive more of \(B\). With threshold \(\theta\), step size \(\Delta r\), and
allowed interval \([r_{\min},r_{\max}]\), the proposed next split is
\begin{equation}
\widetilde r_{k+1}=
\begin{cases}
\min(r_k+\Delta r,r_{\max}), & E_k>\theta,\\[2pt]
\max(r_k-\Delta r,r_{\min}), & E_k<-\theta,\\[2pt]
r_k, & |E_k|\leq\theta.
\end{cases}
\label{eq:dynamic-r-proposal}
\end{equation}
The threshold gives hysteresis: the controller changes \(r\) only when the
evidence is large enough to overcome block-level noise. To avoid using a
predefined cooling schedule, the controller also derives a temperature from
observed hit-rate performance. The block hit rate is
\begin{equation}
h_k=\frac{R_k}{L_k},
\label{eq:dynamic-r-block-hit-rate}
\end{equation}
and the smoothed hit rate is
\begin{equation}
\overline h_k=\alpha h_k+(1-\alpha)\overline h_{k-1}.
\label{eq:dynamic-r-performance-ewma}
\end{equation}
The controller uses \(T_{\max}\) when \(\overline h_k<\eta\) and \(T_{\min}\)
otherwise. For an evidence-supported proposal, the acceptance probability is
\begin{equation}
p_k=
\frac{1}{1+\exp\left(-\dfrac{|E_k|-\theta}{T_k}\right)} .
\label{eq:dynamic-r-acceptance}
\end{equation}
A random draw \(u_k\sim U(0,1)\) accepts the proposed split only when
\(|E_k|>\theta\) and \(u_k<p_k\); otherwise \(r\) is retained. Randomness
therefore affects only whether to accept an evidence-supported proposal, not the
direction of the change.

\subsection{Causality and Outcomes}
\label{sec:dynamic-r-causality}

The adaptation rule is causal. The decision for block \(k+1\) uses only
requests, hits, misses, admissions, promotions, evictions, metadata-history
matches, and queue state observed through block \(k\). It does not use future
requests, next-access times, reuse distances, trace phase labels, offline
selection by workload identity, or counterfactual cache states. The rule also
does not change the capacity budget: \(B\) is fixed during a run, while
Eq.~\eqref{eq:dynamic-r-segment-split} reallocates Cache Capacity between the
two segments.

The experiments below use one representative segmented policy to instantiate
the required causal signals: a controllable split between \(C\) and \(H\), a
metadata-only history signal for objects that were removed from \(C\), and
past-only counters for admissions, promotions, and hits. This choice is an
evaluation instance of the update rule, not a restriction of the method.

Across all three META traces, dynamic \(r\) has little effect at very low
targets, where the minimum cache is already small, but it reduces the required
Cache Capacity at higher targets while maintaining the same hit-rate objective.
The reductions are most meaningful in the regimes where the SLO is stringent
enough that capacity differences translate into material provisioning cost.
These results demonstrate the value of adapting the segment ratio online; they
should be read as evidence for the general causal rule rather than as a
limitation to one policy.

Fig.~\ref{fig:dynamic-r-required-capacity-eta060} gives the corresponding
high-target comparison across all three META traces. In these experiments, Cache
Capacity is counted as the number of objects rather than bytes; the
variable-length object case is discussed separately in
Section~\ref{sec:practical-provisioning}. On META Trace~2, the dynamic controller saves 90,747 cache objects while achieving the same target hit rate of $\eta=0.60$.
Using the measured average object size for META Trace 2, approximately
21.8 MB, this object-count reduction is roughly two terabytes under the
trace-level physical-scale estimate.

\section{Experimental Results}
\label{sec:experimental-results}

\subsection{META Trace 1}
\label{subsec:non-causal-min-fixed}

We next compare replacement policies by the minimum fixed Cache Capacity
\(B^\star\) needed to reach each target hit rate on META Trace 1. This is a
non-causal baseline because \(B^\star\) is obtained by repeated full-trace
replay and binary search. Its purpose is not deployment; it is a common
reference for asking which policy satisfies the same SLO with less Cache
Capacity.


\begin{figure*}[t]
    \centering
    \includegraphics[width=\textwidth]
    {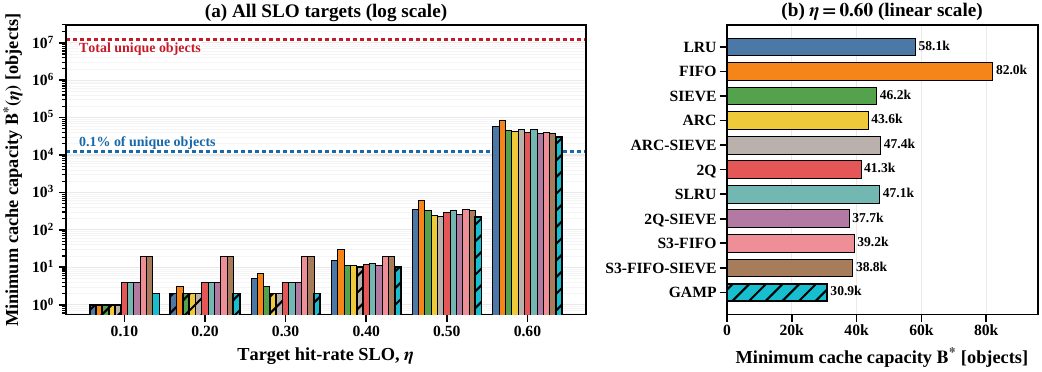}
    \caption{Minimum fixed cache capacity on META Trace 1.
    Panel~(a) reports \(B^\star(\eta)\) for all evaluated policies across
    the target object hit-rate SLOs using a logarithmic scale. The horizontal
    reference lines indicate the total number of unique objects in the trace
    and \(0.1\%\) of that total. Panel~(b) shows the cache-capacity comparison
    at \(\eta=0.60\) using a linear scale. Hatched bars identify the policy
    requiring the minimum cache capacity at each SLO.}
    \label{fig:meta-trace-1-gamp-minfixed-capacity}
\end{figure*}

\begin{figure}[t]
\centering
\includegraphics[width=0.98\linewidth]
{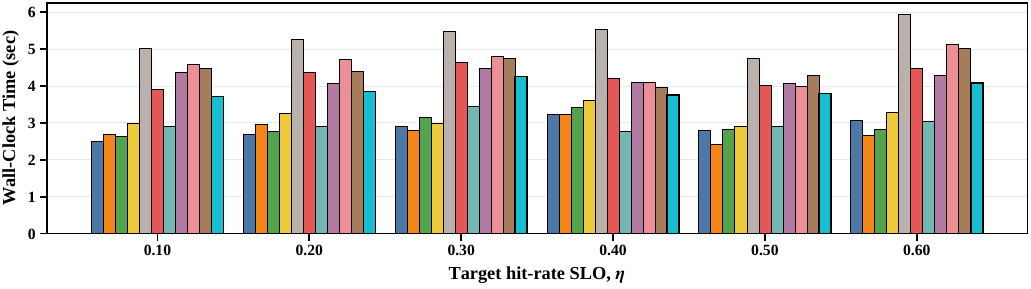}
\caption{GAMP execution-overhead comparison on META Trace 1. The figure reports
wall-clock time at the selected minimum fixed Cache Capacity values. Since all
policies process the same request trace and therefore the same number of
requests, total wall-clock time provides a fair comparison of computational
overhead; equivalently, average processing time per request is obtained by
dividing each reported wall-clock time by the common request count. Across the
shown targets, GAMP has lower overhead than ARC-SIEVE, 2Q, 2Q-SIEVE, S3-FIFO,
and S3-FIFO-SIEVE.}
\label{fig:meta-trace-1-gamp-minfixed-overhead}
\end{figure}

Across target hit rates, no single pre-GAMP baseline is the clear winner: simple
policies are strongest at some low-SLO points, while segmented and SIEVE-based
policies become more competitive as the target increases. This cross-target
variation motivates GAMP's use of a SIEVE-managed hot segment rather than
relying on one fixed existing policy. Figs.~\ref{fig:meta-trace-1-gamp-minfixed-capacity}
and~\ref{fig:meta-trace-1-gamp-minfixed-overhead} add GAMP to the comparison:
at the higher target hit rates, GAMP requires the lowest minimum fixed Cache
Capacity among the evaluated policies, and its wall-clock overhead remains below
ARC-SIEVE, 2Q, 2Q-SIEVE, S3-FIFO, and S3-FIFO-SIEVE. Because every policy
processes the same request trace, the reported total wall-clock times compare
computational overhead under a common request count; per-request processing time
follows by dividing by that same count. The key point of
Fig.~\ref{fig:meta-trace-1-gamp-minfixed-overhead} is therefore not merely that
one policy has a smaller \(B^\star\): the capacity advantage is achieved without
making the implementation the slowest option in the comparison.

\subsection{META Trace 2}
\label{subsec:meta-trace-2}

\begin{figure}[t]
\centering
\includegraphics[width=0.98\linewidth]
{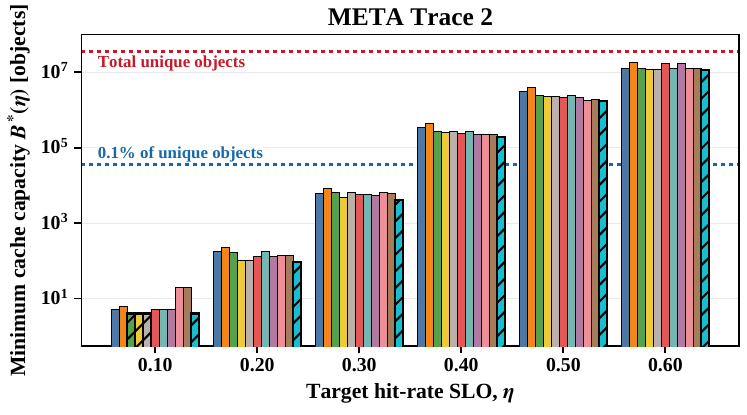}
\caption{Minimum fixed Cache Capacity on META Trace 2. Each bar reports
\(B^\star(\eta)\), counted in objects, for one evaluated policy at a target
object hit-rate SLO. The vertical axis is logarithmic so that low- and
high-target regimes can be compared in the same panel.}
\label{fig:meta-trace-2-gamp-minfixed-capacity}
\end{figure}

Fig.~\ref{fig:meta-trace-2-gamp-minfixed-capacity} shows a much harder
provisioning regime on META Trace 2 than on META Trace 1. The measured
\(B^\star(\eta)\) values rise by several orders of magnitude as the target
increases, and the curve crosses the \(0.1\%\)-of-unique-objects reference
between \(\eta=0.30\) and \(\eta=0.40\). Thus, high hit-rate SLOs require
caching a large resident object set on this trace. This behavior is consistent
with a flatter popularity distribution. Within this measured provisioning task,
GAMP is tied with the best policy at the lowest target and then requires the
least Cache Capacity for the remaining targets.

\subsection{META Trace 3}
\label{subsec:meta-trace-3}

\begin{figure}[t]
\centering
\includegraphics[width=0.98\linewidth]
{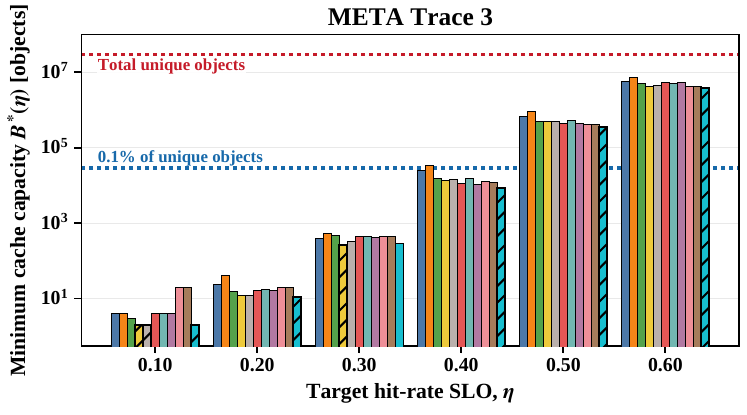}
\caption{Minimum fixed Cache Capacity on META Trace 3. Each bar reports
\(B^\star(\eta)\), counted in objects, for one evaluated policy at a target
object hit-rate SLO. The logarithmic vertical axis emphasizes how quickly
capacity requirements grow as the target hit rate increases.}
\label{fig:meta-trace-3-gamp-minfixed-capacity}
\end{figure}

Fig.~\ref{fig:meta-trace-3-gamp-minfixed-capacity} provides a second high-scale
META-trace case. The low-SLO region again has small absolute capacity
requirements, so the policy differences are operationally limited there. Once the
target reaches the mid-to-high range, the bars move from tens or hundreds of
objects to thousands, then millions of objects. This should be interpreted as an
SLO sensitivity result: a small increase in the required hit rate can imply a
large increase in provisioned cache capacity. The figure does not by itself
identify the cause of that sensitivity, but it shows where policy choice becomes
economically important. In that regime, GAMP remains among the lowest-capacity
choices, including the best result at the higher targets.




\subsection{Synthetic Traces}
\label{subsec:synthetic-traces}

\begin{figure}[t]
    \centering
    \includegraphics[width=0.98\linewidth]
    {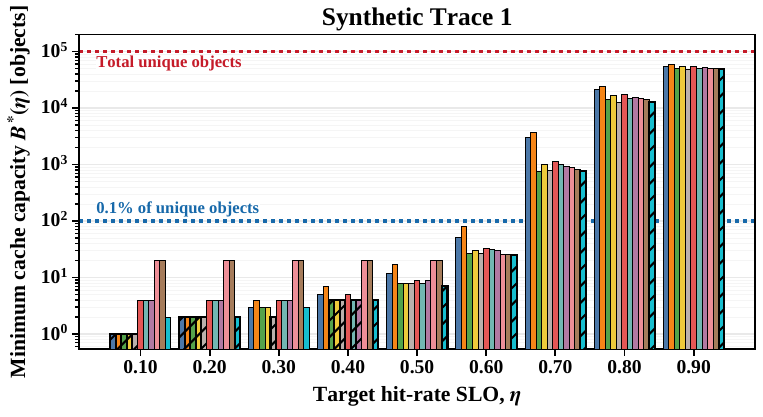}
    \caption{Minimum fixed cache capacity on Synthetic Trace~1. Each bar
    reports \(B^\star(\eta)\), counted in objects, for one evaluated policy
    at a target object hit-rate SLO. The synthetic trace includes higher
    target values, \(\eta=0.70\), \(0.80\), and \(0.90\), to stress the
    policies under stricter provisioning requirements.}
    \label{fig:synthetic-gamp-minfixed-capacity}
\end{figure}

Fig.~\ref{fig:synthetic-gamp-minfixed-capacity} complements the META traces with
a controlled non-stationary workload. At low targets, the required cache sizes
are close to the bottom of the tested range, so policy differences are small in
absolute terms. Once the target exceeds \(\eta=0.60\), the required capacity
increases sharply, and the policy choice again becomes a provisioning decision
rather than a minor implementation detail. GAMP remains in the low-capacity group
across this harder regime, supporting the same conclusion observed on the META
traces. The histogram for the second synthetic trace is very similar to Fig.~\ref{fig:synthetic-gamp-minfixed-capacity}. Therefore, we omit the standalone plot and instead include both synthetic traces in Fig.~\ref{fig:cross-trace-eta50} below. Despite the tenfold difference in catalog size, the two synthetic traces exhibit similar relative policy performance.

\subsection{Cross-Trace Comparison}
\label{subsec:cross-trace-comparison}

\begin{figure}[t]
    \centering
    \includegraphics[width=0.98\linewidth]
    {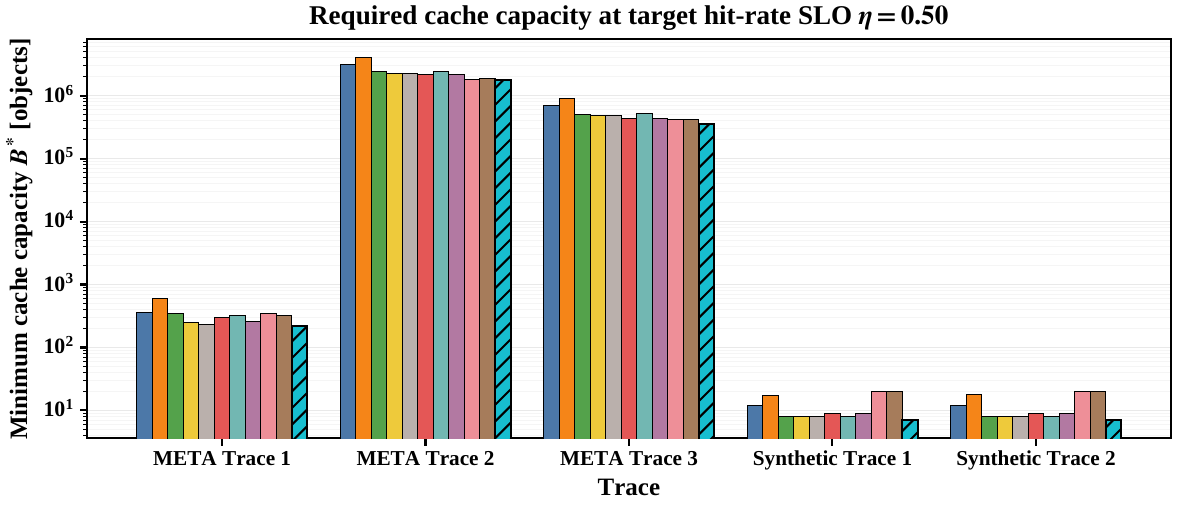}
    \caption{Cross-trace comparison of the minimum cache capacity
    \(B^\star\) required to satisfy the target hit-rate SLO
    \(\eta=0.50\). Each group corresponds to one of the five workloads,
    and the bars represent the eleven evaluated caching policies. The
    vertical axis is logarithmic because the required capacities differ
    by several orders of magnitude across the workloads. Policy colors
    are consistent with the preceding figures, and hatched bars identify the policy requiring the minimum cache capacity. }
    \label{fig:cross-trace-eta50}
\end{figure}

Fig.~\ref{fig:cross-trace-eta50} compares the provisioning requirements of
the five workloads under the same target hit-rate SLO. The large variation
in \(B^\star\) reflects differences in request concentration and effective
working-set size. When object popularity is highly concentrated, a relatively
small cache can retain the frequently requested objects and achieve the target
hit rate. In contrast, a flatter popularity distribution spreads requests
across a larger working set and therefore requires substantially more cache
capacity to satisfy the same SLO. Accordingly, META Trace~2 requires the
largest cache capacities, followed by META Trace~3, whereas META Trace~1 and
the two synthetic traces require considerably smaller capacities at
\(\eta=0.50\). Notably, although Synthetic Trace~2 contains ten times as many
objects as Synthetic Trace~1, their required capacities at this target are
nearly identical, indicating that catalog size alone does not determine
\(B^\star\); the concentration of requests within the catalog is also
critical. Despite these workload-dependent differences, GAMP requires the
minimum \(B^\star\) on all five traces, indicating that its capacity advantage
persists across workloads with substantially different provisioning
requirements.

\subsection{Traditional Fixed-Capacity Comparison}
\label{subsec:traditional-cache-comparison}

The preceding experiments evaluate cache provisioning by determining the
minimum capacity \(B^\star\) required to satisfy a target hit rate. We
complement that analysis with the traditional fixed-capacity comparison, in
which every policy receives the same cache capacity \(B\) and is evaluated by
the resulting object hit rate. This comparison represents deployments in
which the cache budget has already been selected, for example because of
hardware, memory, or operational constraints, and the remaining question is
which policy uses that fixed capacity most effectively.

\begin{figure*}[t]
    \centering
    \includegraphics[width=\textwidth]
    {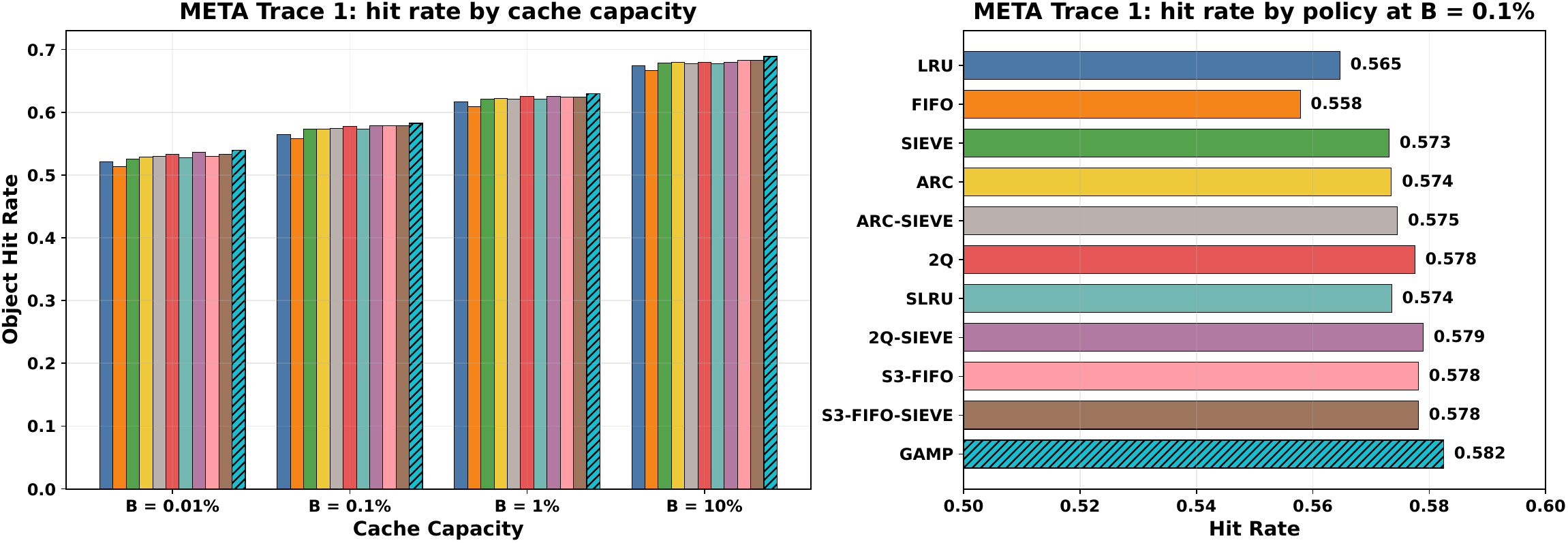}
    \caption{Traditional fixed-cache-capacity comparison on META Trace~1.
    The left panel reports the object hit rate of each policy when all policies
    are assigned the same cache capacity,
    \(B\in\{0.01\%,0.1\%,1\%,10\%\}\), expressed as a percentage of the
    distinct-object population. The right panel enlarges the comparison at
    \(B=0.1\%\), where the small differences among policies are more readily
    visible. In each comparison, hatching identifies the policy with the
    highest observed hit rate. GAMP achieves the highest hit rate at every
    evaluated capacity, showing that its minimum-capacity advantage is
    consistent with the conventional fixed-budget evaluation.}
    \label{fig:meta-trace-1-traditional-hit-rate}
\end{figure*}

Fig.~\ref{fig:meta-trace-1-traditional-hit-rate} reports this comparison for
META Trace~1. The left panel evaluates all eleven policies at cache capacities
equal to \(0.01\%\), \(0.1\%\), \(1\%\), and \(10\%\) of the distinct-object
population. As expected, the hit rate increases with cache capacity for every
policy. However, GAMP achieves the highest observed hit rate at each evaluated
capacity, as indicated by the hatched bars. Thus, GAMP's advantage in the
minimum-\(B^\star\) experiments is not solely a consequence of the capacity
search procedure: it also remains the strongest policy when all policies are
given an identical cache budget.

Because the differences among several policies are difficult to distinguish
in the grouped capacity sweep, the right panel provides an expanded
policy-by-policy view at \(B=0.1\%\). At this commonly used fixed-capacity
operating point, GAMP again achieves the highest object hit rate. This
traditional comparison therefore complements the minimum-capacity results:
the latter measures provisioning efficiency for a target performance level,
whereas Fig.~\ref{fig:meta-trace-1-traditional-hit-rate} measures the
performance obtained from an already provisioned cache. We also performed the same fixed-capacity comparison on META Traces~2 and~3 and both synthetic traces for \(B \in \{0.01\%, 0.1\%, 1\%, 10\%\}\). For each evaluated workload, GAMP attains the highest hit rate at all fixed-capacity settings.

\begin{figure}[t]
    \centering
    \includegraphics[width=0.90\linewidth]
    {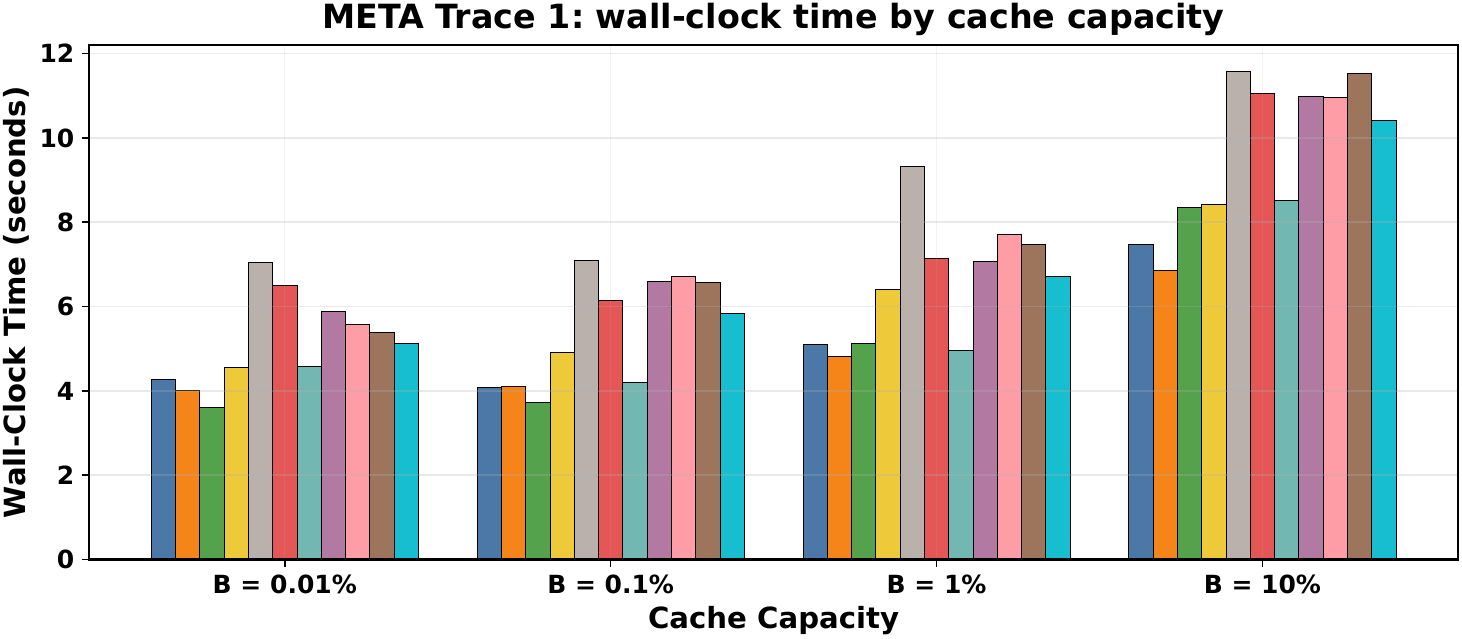}
    \caption{Wall-clock time for the traditional fixed-cache-capacity
    comparison on META Trace~1. Each policy processes the complete trace at
    the same cache capacities,
    \(B\in\{0.01\%,0.1\%,1\%,10\%\}\), expressed as percentages of the
    distinct-object population. Policies follow the same left-to-right order
    and color mapping as in
    Fig.~\ref{fig:meta-trace-1-traditional-hit-rate}. Wall-clock time measures
    the total execution time for processing the trace and serves as a proxy
    for computational overhead.}
    \label{fig:meta-trace-1-traditional-wall-clock}
\end{figure}

\section{Discussion: Practical Capacity Provisioning and Pricing at the Edge}
\label{sec:practical-provisioning}

\begin{center}
    \refstepcounter{table}\label{tab:edge-resource-models}
    {\footnotesize\textbf{TABLE~\Roman{table}}\\
    Representative edge and near-edge resource prices. Exact availability and
    prices may depend on region and service configuration.}
    \vspace{0.6ex}
    \scriptsize
    \setlength{\tabcolsep}{2pt}
    \renewcommand{\arraystretch}{1.08}
    \begin{tabular}{@{}p{0.27\columnwidth}p{0.49\columnwidth}p{0.17\columnwidth}@{}}
        \toprule
        Service & Example allocation and price & Billing unit \\
        \midrule
        AWS Wavelength EC2
        &
        4 GB: \$0.056/hour; 16 GB: \$0.224/hour
        &
        instance-hour~\cite{AWSWavelengthPricing}
        \\
        AWS Wavelength EBS
        &
        provisioned volume: about \$0.15/GB-month
        &
        GB-month~\cite{AWSWavelengthPricing}
        \\
        Akamai shared CPU
        &
        1 GB RAM, 1 vCPU, 25 GB SSD: \$5/month or \$0.0075/hour
        &
        VM-month/hour~\cite{AkamaiComputePlans}
        \\
        Akamai dedicated CPU
        &
        4 GB RAM, 2 vCPUs, 40 GB SSD: \$36/month or \$0.05/hour
        &
        VM-month/hour~\cite{AkamaiComputePlans}
        \\
        Akamai high memory
        &
        24 GB RAM, 2 vCPUs, 20 GB SSD: \$60/month or \$0.09/hour
        &
        VM-month/hour~\cite{AkamaiComputePlans}
        \\
        Cloudflare Workers
        &
        128 MB/isolate; 30M CPU-ms included; \$0.02/million CPU-ms after that
        &
        CPU-ms~\cite{CloudflareWorkersLimits}
        \\
        Cloudflare Workers KV
        &
        1 GB stored data included; \$0.50/GB-month after that
        &
        GB-month + ops~\cite{CloudflareKVPricing}
        \\
        \bottomrule
    \end{tabular}
\end{center}

The above evaluations report resident Cache Capacity \(B\) in objects. With
variable-length objects, physical capacity is measured in bytes and depends on
the  object lengths. Therefore, translating object-count reductions to
byte-scale savings is approximate and depends on the workload (trace).

The mapping from physical capacity to monetary cost is also
deployment-specific. Edge and near-edge platforms expose different resource
models, including virtual machines with bundled RAM, compute, and local
storage; serverless execution environments with fixed per-isolate memory
limits; and managed storage services billed according to stored data and
operations. Table~\ref{tab:edge-resource-models} gives representative public
pricing examples. These offerings are not directly interchangeable; rather,
they illustrate that the resource represented by \(B\) may be provisioned and
billed differently across deployment architectures.

In particular, reducing \(B\) does not necessarily reduce the provider bill
proportionally: the deployment may remain in the same tier, although memory or
storage is freed. Direct monetary saving occurs when the reduction permits a
smaller tier, removes an instance or replica, or decreases usage-metered
storage. Using the two-terabyte saving discussed in
Section~\ref{sec:dynamic-r-causality}, the AWS Wavelength EBS example price in
Table~\ref{tab:edge-resource-models} gives about
\(2 \times 1024 \times \$0.15 \approx \$307\) per cache replica per month,
before replication, operations, and regional price variation. Thus, \(P=B\)
is a provider-independent resource-cost proxy; an exact monetary function
\(P(B,E)\) depends on the deployment platform, resource type, replication
level, and pricing tier.
Nevertheless, a memory resource that is \emph{shared} by different user services forming a system may overall perform better if the cache process occupies less memory, thus making more memory available to the other services, even if this does not result in a cost-savings (recall the related discussion in  Section \ref{sec:introduction}).

\section{Conclusions and Future Work}
\label{sec:conclusions-future-work}
For different operator/user-specified hit rates, we evaluated
several pure and segmented policies in terms of their cost,
particularly their required capacity (amount of memory),
but also their computational complexity. This approach 
better aligns with how users/operators engage the cloud, 
particularly the edge or near-edge cloud for the context of
a content cache. We also proposed 
a hybrid-segmented policy GAMP and a causal segment-ratio adaptation rule for
responding to the non-stationary most-popular (i.e., working) set of objects.
The performance study was conducted on three qualitatively different,
week-long benchmark META traces and two controlled non-stationary
synthetic traces.
GAMP was shown to have the least cache capacity requirement across wide
ranges of user-specified hit-rates and query workloads, with competitive
computational overhead per query.

The foregoing policies are readily adapted to the case of variable-length objects where cache capacity will thus be measured in a unit/quanta
of memory, e.g., megabytes. (Recall the footnote at the end of Section \ref{sec:workloads}.)
One basic approach is to scale-down  by 
its length $\ell$ the eviction-rank $\rho$ 
of a cached object, i.e., $\rho/\ell$. Thus, if there are two objects of
very different lengths but nearly equal rank, the longer object is evicted first.
In future work, we will produce a similar performance study for such policies.

In future work, we will also produce a 
performance study wherein the cache capacity is also dynamically
adapted. Here we naturally expect the time-average
of a good approach will be slightly larger than the ``non-causal" capacity minima
reported above, but the variance will, however, depend
on the dynamism of the 
working set.

\end{document}